# Enhanced betatron radiation by steering a low-energy-spread electron beam in a deflected laser-driven plasma wiggler

Changhai Yu[1,2], Jiansheng Liu[1,3,⋆], Wentao Wang[1], Wentao Li[1], Rong Qi[1], Zhijun Zhang[1], Zhiyong Qin[1,2], Jiaqi Liu[1,2], Ming Fang[1,2], Ke Feng[1,2], Ying Wu[1,2], Cheng Wang[1], Yi Xu[1], Yuxin Leng[1,3], Changquan Xia[4], Ruxin Li[1,3,5,⋆], and Zhizhan Xu[1,3,5,⋆]

[1]State Key Laboratory of High Field Laser Physics, Shanghai Institute of Optics and Fine Mechanics, Chinese Academy of Sciences, Shanghai 201800, China

[2]University of Chinese Academy of Sciences, Beijing 100049, People's Republic of China

[3]IFSA Collaborative Innovation Center, Shanghai Jiao Tong University, Shanghai 200240, China.

[4]College of Physical Science and Technology, Yangzhou University, Jiangsu 225001, China

[5]School of Physical Science and Technology, ShanghaiTech University, Shanghai 20031, China

**Laser wakefield accelerators (LWFA) hold great potential to produce high-quality high-energy electron beams (*e* beams) and simultaneously bright x-ray sources via betatron radiation, which are very promising for pump-probe study in ultrafast science. However, in order to obtain a high-quality *e* beam, electron injection and acceleration should be carefully manipulated, where a large oscillation amplitude has to be avoided and thus the emitted x-ray yield is limited. Here, we report a new scheme to experimentally enhance betatron radiation significantly both in photon yield and photon energy by separating electron injection and acceleration from manipulation of the *e*-beam transverse oscillation in the wake via introducing a slanted thin plasma refraction slab. Particle-in-cell simulations indicate that the *e*-beam transverse oscillation amplitude can be increased by more than 10 folds, after being steered into the deflected laser-driven wakefield due to refraction at the slab's boundaries. Spectral broadening of the x-rays can be suppressed owing to the small variation in the peak energy of the low-energy-spread *e* beam in a plasma wiggler regime. We demonstrate that the high-quality *e*-beam generation, refracting and wiggling can act as a whole to realize the concurrence of monoenergetic *e* beam and bright x-rays in a compact**

**LWFA.**

X-ray synchrotron radiation sources have become immensely useful tools for basic science and broad applications in biology, and material science[1]. State-of-the-art synchrotrons and free-electron lasers[2,3] based on a radiofrequency accelerator can now produce x-ray sources with unprecedented photon flux and brilliance, but have hitherto been limited to huge facilities which are costly and only accessible to limited users. Over the past decade, a more compact accelerator based on the concept of laser-driven wakefield acceleration[4] has achieved significant progress in generating GeV-class electron beams (*e* beams)[5-11], which holds great potential of becoming a better candidate to produce compact femtosecond x- and γ-ray sources[12]. In such a laser wakefield accelerator (LWFA), electrons in the wakefield would witness an ultrahigh longitudinal acceleration field above 100 GV/m, simultaneously undergo a betatron oscillation as a result of the transverse focusing field of the wake and emit bright high-energy x-rays over a few millimeters through the betatron radiation mechanism[13-15]. Attributed to the intrinsic synchronization to the driving laser pulse, both the generated *e* beam and the x-ray pulse from the same wakefield have an ultrashort duration of several femtoseconds, which are very promising for pump-probe study in ultrafast science.

The properties of betatron radiation are normally characterized by the strength parameter[14-16] $K = \gamma \omega_\beta r_\beta / c \simeq 1.33 \times 10^{-10} \sqrt{\gamma n_e \left[ cm^{-3} \right]} \cdot r_\beta \left[ \mu m \right]$, where $r_\beta$ is the amplitude of betatron oscillation, γ is the relativistic factor of the electron, $n_e$ is the plasma density, $\omega_p$ and $\omega_\beta = \omega_p / (2\gamma)^{1/2}$ are the plasma and betatron frequencies. Betatron radiation spectrum is peaked at the fundamental frequency in the undulator regime for $K \leq 1$. However, in the wiggler regime for $K \gg 1$, it is broadened consisting of merged harmonics characterized by the critical frequency $\hbar\omega_c \approx 5.24\times10^{-24}\gamma^2 n_e\left[cm^{-3}\right]r_\beta\left[\mu m\right]$. The average number of photons emitted by the $e$ beam is given by, $\langle\overline{N}_X\rangle \simeq (2\pi/9)(e^2/\hbar c)N_0 N_e K \simeq 5.6\times10^{-3}N_0 N_e K$, where $N_0$ is the number of betatron oscillations executed by the $e$ beam, and $N_e$ is the number of wiggling electrons[17]. Therefore, an effective way to increase the photon energy and yield of betatron radiation is to increase the oscillation amplitude $r_\beta$ in additional to increasing the $e$-beam charge and energy. However, in order to obtain a high-quality high-energy $e$ beam, electron injection and acceleration should be carefully manipulated via a well-performed LWFA, where a large oscillation amplitude $r_\beta$ has to be avoided. Some ideas of manipulating $r_\beta$ to enhance betatron radiation have been demonstrated, e.g. by enhancing betatron oscillations resonantly in a plasma wake[18], applying the driving pulse-front-tilt to induce off-axis electron injection[19], appropriately tailoring the plasma profile[20], using a clustering gas jet[21] or via an evolving laser-plasma bubble[22]. However, the produced x-rays had continuum spectra because the controllability was limited in the $e$-beam quality. Recently, an idea of a helical plasma undulator has been proposed to produce controllable synchrotron-like radiation by inducing centroid oscillations of the laser pulse[23, 24] in a plasma channel.

In this article, we have experimentally realized a new scheme to enhance the betatron radiation via separating electron injection and acceleration from manipulation of the $e$-beam transverse oscillation in the wakefield. By producing a 30-μm-thick slanted plasma slab (SPS) with a higher density somewhere in the acceleration stage, the driving laser pulse can be deflected owing to the refraction, but the accelerated monoenergetic $e$ beam can almost keep its initial propagation axis and obtain an increased transverse momentum instantly due to the axes misalignment. By this way, the high-quality $e$ beam can be steered into the deflected laser-driven wakefield with a controllable operation both in the transverse oscillation amplitude and energy of the *e* beam. Furthermore, by restraining the fluctuation of the *e*-beam energy in the wakefield, spectral broadening of betatron can be suppressed in a plasma wiggler regime. Brilliant betatron x-rays (~$10^{23}$ photons s$^{-1}$ mm$^{-2}$ mrad$^{-2}$ 0.1% BW) in tens of keV have been produced with significant enhancement both in photon yield and peak energy.

**Results**

**Electron-beam generation and manipulation.** Figure 1 shows the experimental setup. The experiments were carried out at the femtosecond 200TW laser system[25]. The laser pulses with an on-target power of 100 TW were focused to reach a peak intensity of $3.6 \times 10^{18}$ W/cm$^2$ (see Methods). The laser beam distribution at the focus was also optimized to have a smooth profile to avoid laser filamentation[26, 27]. A LWFA consisting of two-segment pure helium gas jets was designed to generate high-quality *e* beams with low energy spread, sufficient charge, and low divergence[9, 28, 29]. An optical interferometer and a shadowgraphy were set up to measure the plasma density distribution as shown in Fig. 1b (see Methods). It was seen that a thin SPS with a thickness of ~30 μm and a skew angle of ~$30^0$ was produced (see Methods). The plasma densities $n_\text{I}$ in front of the slab, $n_\text{II}$ within the plasma slab and $n_\text{III}$ behind the slab were measured[30, 31] to be $(1\pm0.1)\times10^{19}$ cm$^{-3}$, $(5\pm2)\times10^{19}$ cm$^{-3}$ and $(6\pm0.5)\times10^{18}$ cm$^{-3}$, respectively. This SPS was operated as a refraction plate to deflect the driving laser and thus to manipulate the *e*-beam transverse oscillation in the wakefield.

The physical scenario of using the SPS to steer the transverse oscillation of electrons in a laser-driven wakefield was depicted in Fig. 2. The refractive indices for three-segment plasmas can be calculated by $\eta = \left[1 - \omega_p^2 / \gamma_\perp \omega^2\right]^{1/2} \approx 1 - n_e / 2\gamma_\perp n_c$, where $\omega$ is the laser frequency, $\gamma_\perp$ is the relativistic factor depending on the laser intensity, and $n_c$ is the critical density[32, 33]. When the laser pulse entered the front boundary of the slab, it was deflected off the incident direction with a deviation angle of $\Delta\theta_{I \to II} \approx -5.0$ mrad and then with $\Delta\theta_{II \to III} \approx 5.6$ mrad at the rear boundary, abiding by the Snell's law $\eta_I \sin\theta_I = \eta_{II} \sin\theta_{II} = \eta_{III} \sin\theta_{III}$. The wakefield was deflected in the same way as the driving laser, the accelerated $e$ beam would thus gain an extra transverse momentum instantly due to the axes misalignment of the laser and the $e$ beam[24]. Then the electrons undergoing the transverse focusing of the wakefield and bending[34, 35] at the boundaries would be steered from its initial propagation. Finally, the $e$ beam generated in front of the SPS would obtain a much larger transverse oscillation amplitude in the plasma wiggler behind the SPS, where the $e$-beam energy would not change much if the $e$ beam was located close to the zero-phase region of the wakefield.

Electron generation was investigated firstly by comparing the two cases when the SPS was introduced or not. As shown in Fig. 3a, only one monoenergetic $e$ beam with a clean background and a small divergence of ~ 0.2 mrad was measured. However, once introducing the slab, the $e$ beams would have a larger divergence of ~ 1 mrad and the peak energy varied from 259 to 351MeV while shifting the position of the slab from z =1.3 to 2 mm, as shown in Fig. 3b-d. In addition to these monoenergetic $e$ beams, some lower-energy electrons tails were also observed, which might be attributed to shock-front injection at the rear of the slab[36]. The introduced SPS did not deteriorate the energy spread of $e$ beams and the integrated charge around the peak energy kept roughly unchanged with the similar uncertainty (Fig. 3f). Moreover, the statistic fluctuation of $e$-beam central positions recorded at 3.6 m away from the gas jet was estimated to be less than 1mrad both for the two cases (see Fig. 3g). This fluctuation mainly came from the shot-to-shot fluctuation in the jitter of the laser and density distribution. However, the averaged central position shifted upward with a deflection of ~ 0.6 mrad while introducing the SPS, indicating that the refraction of the driving laser by the SPS affected the *e*-beam transverse oscillation.

**Numerical modeling.** Particle-in-cell (PIC) simulations using the Vorpal Code were carried out to get an insight into the details of the *e*-beam generation and manipulation (see Methods). Without the SPS, one high-quality *e* beam can be accelerated up to a peak energy of 465 MeV, with the energy spread of 3.2% in full width at half maximum (FWHM), 0.6 mm mrad normalized emittance, and 18 pC charge (Fig. 4c). The injected electrons keep being accelerated well longitudinally from z=0.96 to 3.6 mm with a transverse oscillation amplitude smaller than 0.18 μm. However, while introducing the SPS at z=1.5 mm as shown in Fig. 4a and keeping other parameters the same, the deflection of the laser-driven wakefield in the vertical direction was observed in Fig. 4b. Due to the refraction, the propagation direction of the wakefield is deflected from its initial direction in addition to an absolute offset in the vertical direction. After being steered into the deflected wakfield, the accelerated *e*-beam transverse radius $R_b$ increases rapidly from 0.18 to 1.9 μm, as shown in Fig. 4b and 4d. The *e* beam is also deflected from the original laser propagation direction with a deviation angle of 0.7 mrad, in good agreement with the aforementioned measurement and analysis.

Besides, the *e*-beam length remains the same without loss of beam charge and the relative energy spread can even be reduced a little, which might be attributed to energy chirp compensation due to the phase space rotation[37, 38]. Some electrons might be injected into the wakefield behind the slab owing to the shock-front injection at the transient downward density ramp[29], but they can't be efficiently accelerated due to the quick dephasing, because there is a great density difference at the downward density ramp. However, the steered *e* beam slips forward quickly with respect to the wakefield to the zero-phase region and thus the *e*-beam energy varies little in the following acceleration stage, which is meant to be operated as a plasma wiggler. Furthermore, the FWHM energy spread of the *e* beams in the peak produced in our case is as small as 2.5%. These two effects and periodical oscillation in the wiggler stage are expected to reduce the bandwidth of the betatron radiation spectrum near the peak, which is supported by the simulated radiation distribution from the Lienard-Wiechert potentials according to electrons trajectories. In this scheme, the high-quality *e*-beam generation, refracting and wiggling can act as a whole to realize the concurrence of monoenergetic *e* beam and bright x-rays in a compact LWFA by manipulating the transverse oscillation of the *e* beam .

**Enhanced betatron x-ray radiation.** Two techniques were employed to measure the betatron radiation spectra in a single-shot (see Methods). Firstly, in the case of no SPS when the yield of betatron radiation was low, an x-ray CCD camera operated in a single-photon-counting (SPC) mode was used to measure the spectra of the betatron radiation, which could also be used to measure the *e*-beam transverse size and emittance as well [39-41]. Secondly, since the x-ray emission would be enhanced both in photon yield and photon energy after introducing the SPS, the SPC technique was not suitable any more for a much higher photon flux and higher photon energies. Then an x-ray detection system (XRDS) based on the x-ray transmission through an array of filters[28, 42, 43] was designed to measure the radiation spectra (see Methods). In order to avoid the influence of the driving laser on the detector, a 8-µm-thick Al film or 300-µm-thick Be window was placed in front of the CCD camera or the LSO-crystal scintillator to block the laser.

In the case of no SPS corresponding to the *e*-beam generation at 465 MeV in Fig. 3a, the recorded betatron radiation pattern via the XRDS, the recorded radiation signal on the x-ray CCD chip, and retrieved radiation spectrum via the SPC were shown respectively in Fig. 5a-c. According to the retrieved spectral profile (Fig. 5c), the critical photon energy was estimated to be 5.8±0.4 keV, indicating that the betatron oscillation amplitude $r_\beta$ was less than 0.2 μm. This was slightly larger than a matched e-beam size of ~0.1 μm given by the theoretical prediction $\sigma_x = \sigma_\theta \left( \lambda_p / \pi \right) \sqrt{\gamma / 2}$, where $\sigma_\theta$ was the e-beam divergence[4, 13, 41]. The undulator strength parameter was thus calculated as $K = \gamma \omega_\beta r_\beta / c \simeq 2.1$ and the number of betatron oscillation was estimated as $N_0 = L/\lambda_b \approx 5$, where $L$ was the length of the wiggler, $\lambda_b = \sqrt{2\gamma} \lambda_p$ was the betatron wavelength. The total photon yield could be predicted to be ~$1.5 \times 10^7$. From the recorded betatron radiation pattern on the XRDS, the photon number of betatron radiation was measured to be $(1.1 \pm 0.2) \times 10^7$, in a reasonable agreement with the predicted one. The detected x-ray beam had divergence angles of 2.8 and 2.2 mrad in horizontal and vertical directions, respectively, consistent with the divergence estimated by $\theta \cong K/\gamma$ ( ~ 2.5 mrad for γ = 835) around the electron velocity vector.

While introducing a SPS in the acceleration stage to manipulate the e-beam transverse oscillation and betatron radiation, the x-ray spectra were measured via the XRDS. Shown in Fig. 5d and 5e were the recorded radiation patterns without and with inserting filters respectively, when the SPS was introduced at z=1.5 mm. As shown in Fig. 3b, the generated e beam in this case had the peak energy of 259 MeV with a FWHM energy spread of 1.8%. The retrieved radiation spectrum was shown in Fig. 5f. It was found the FWHM bandwidth of betatron radiation spectra in the peak was about 55%, although the higher harmonics could have broadened the spectrum due to the strength parameter ($K \approx 18$). In spite of a much lower e-beam energy in this case, the critical energy of the radiated x-ray was increased from 5.5 keV to 26 keV, and the photon number which was estimated to be $(2.1\pm0.8)\times10^8$ was also enhanced by more than 12 folds, as compared with the radiation spectrum (Fig. 5c) for the case of no SPS. Assuming that the size and duration were around 4 μm and 6 $f$s, the x-ray source could have a peak brilliance of ~$10^{23}$ photons s$^{-1}$ mm$^{-2}$ mrad$^{-2}$ 0.1% BW. By varying the SPS position (with z increasing), the $e$-beam energy was increased from 259 to 351 MeV (Fig. 3b-d) and the corresponding critical photon energy of betatron radiation shifted from 22 to 34 keV (Fig. 5f). For each of the aforementioned cases, both the statistical average x-ray photon energy and yield were increased greatly (Fig. 5g). Besides, the enhanced betatron radiation had a larger divergence angle if compared with the case of no SPS (see Fig. 5a,d) and the central position of the radiation was also shifted upward with a deviation angle of ~ 0.8 mrad, roughly corresponding to the deflection of the generated $e$ beam. These results verified that the high-quality $e$ beam could be steered into the deflected laser-driven wakefield with a controllable operation both in the $e$-beam energy and transverse oscillation amplitude via introducing a SPS, as demonstrated in the previous section of numerical modeling.

**Discussion**

We have experimentally realized a new scheme of separating electron injection and acceleration from manipulation of the $e$-beam transverse oscillation. By introducing a thin SPS with a higher density somewhere in the acceleration stage, both the transverse oscillation amplitude and the $e$-beam energy could be manipulated successfully in the wakefield, which was supported by the PIC simulations as well. Previous methods reported to enhance betatron radiation by enhancing betatron oscillations resonantly in a plasma wake[18] or applying the driving pulse-front-tilt to induce off-axis electron injection[19] can not maintain a low-energy-spread $e$ beam. However in this scheme, the produced *e* beam with a low energy spread in a well performed LWFA was steered into the zero-phase region of a deflected wakefield, which was operated as a plasma wiggler. By this way, the *e* beam acquired a large transverse oscillation amplitude but at the same time the *e*-beam energy varies little. Therefore the periodical oscillation and radiation in the plasma wiggler with a relatively stable *e*-beam energy would significantly reduce the bandwidth of the betatron spectrum in the peak. The high-quality *e*-beam generation, refracting and wiggling can act as a whole to realize the concurrence of monoenergetic *e* beam and bright x-rays in a compact LWFA by manipulating the transverse oscillation of the *e* beam. It is anticipated that this compact monoenergetic *e* beam and brilliant x-ray source will provide practical applications in ultrafast pump-probe study.

**Methods**

**Laser system.** The experiments were carried out at the femtosecond 200TW laser system based on the chirped-pulse amplification (CPA) Ti:sapphire with a 1-Hz repetition rate. The 33-*f*s, 800-nm laser pulses with a beam diameter of roughly 80 mm and an on-target power of 100 TW were focused by an *f*/30 off-axis parabola mirror into the gas jet and the vacuum beam radius $w_0$ was measured to be 32 $u$m at $1/e^2$, the peak intensity was estimated to be $3.6 \times 10^{18}$ W/cm$^2$, corresponding to a normalized amplitude of $a_0 = 1.3$. The fractional laser energy contained within the laser spot was measured to be ~ 59%.

**Gas jet and density measurement.** The gas target consisted of two-segment pure (0.8mm+3mm) helium gas jets, which were used to produce a structured gas flow for realizing cascaded acceleration[28, 29]. A probe beam split from the main laser beam was sent perpendicularly across the gas jet, then entered a 4*f* Michelson-type interferometer and shawdowgraphy using a 4*f* optical imaging system for measuring the plasma density. The first-segment gas jet was filled with pure He atoms at a pressure of 5bar, once ionized, this pressure corresponded to an electron background density of $(1\pm0.1)\times10^{19}$cm$^{-3}$ and the second-segment gas flow was operated with an average plasma density of $(6\pm0.5)\times10^{18}$cm$^{-3}$ as measured. A wedge-shaped face was then fabricated at the top edge of the right wall of the first-segment gas cell, which was placed at the left edge of the second gas jet, to produce an inclined shock front[36, 44] by stopping the supersonic gas flow, and thus a slanted thin gas layer was formed with a higher density than the ambient one.

**Electron beam measurements.** The laser-accelerated *e*-beams were deflected by a 90-cm-long tunable dipole electromagnet (3.6 m after the exit of the gas jet) with a maximum magnetic field of 1.1 Tesla, corresponding to an energy resolution of 0.06% at 300 MeV with 0.1 mrad divergence, and measured by a Lanex phosphor screen (PS) imaged onto an intensified charge-couple device 16-bit (ICCD) in a single shot, which was cross-calibrated by using a calibrated imaging plate and an integrating current transformer (ICT) to measure the charge of the *e*-beams.

**PIC Simulations.** The *e*-beam generation in the underdense plasma and manipulation by the slanted plasma slab were all performed using the Vorpal code. A linearly p-polarized Gaussian laser pulse with $\lambda_L$=800 nm, $a_0$=1.6, $w_0$=32 μm (FWHM) and $\tau_{FWHM}$ = 30 *f*s was chosen to match the experimental condition. A moving window with a size of 75×160 μm$^2$ was used. The grid cell size was $k_0z$=0.209 in the laser propagation direction and $k_0x$=0.393 in the transverse direction with four macroparticles per cell. The longitudinal plasma profile consisted of an 192-μm-long upward density ramp followed by a 568-μm-long plateau with a density of 1×10$^{19}$cm$^{-3}$, and then a 160μm-long downward density ramp from the maximum density of 1.7×10$^{19}$cm$^{-3}$ to $0.65\times10^{19} cm^{-3}$, which was followed by a 2.5-mm-long slow downward plasma with an average density of $0.45\times10^{19} cm^{-3}$. A slanted plasma slab with a thickness of 25 μm and a skew angle of $26.5^0$ was then introduced with the varied position in the acceleration stage.

**X-ray beam detection and analysis.** Firstly, a back-illuminated x-ray charge-coupled device (CCD) camera with 1340×400 pixels of size 20×20 μm² operated in a single-photon-counting (SPC) mode[39-41] was used to measure the spectra below 20 keV, which was set at 4.7m from the gas jet and protected from the residual laser-light by an 8-μm-thick Al foil in front of the x-ray CCD. The piling events[45] could then be avoided by satisfying the low photon flux requirement.

Secondly, an x-ray detection system (XRDS) based on the x-ray transmission through an array of filters[28, 42, 43] was also designed to measure the spectrum. The driving laser beam was blocked by a 300-μm-thick Be window. The fluorescence signal of x-ray beams from the $Lu_2SiO_4$(LSO)-crystal scintillator with a diameter of 50 mm was imaged onto a CCD camera within a collecting solid angle of 9.6 mrad, which was placed downstream 5.2 m away from the gas jet. The calibrated XRDS was encased by the lead baffles to minimize the background noise, taking into account the intrinsic detector response including quantum efficiency as well as the transmitted materials. Following the method of Ta Phuoc *et al.*, the spectrum can be derived from the signal difference transmitted through Al, Cu and Pb filters. Assuming a sufficiently narrow form of the function $f_k(\hbar\omega) = (T_k(\hbar\omega) - T_{k+1}(\hbar\omega))R(\hbar\omega)/(10^{-3}\hbar\omega)$, where $T_k$ is the transmittance of filter $k$ and $R$ is the LSO response (conversion from photon number to imaging CCD counts). The number of photons per 0.1% bandwidth, assigned to the mean of the distribution $f_k(\hbar\omega)$, is $\left[10^{-3}\hbar\omega dN_\gamma / d(\hbar\omega)\right]_{\hbar\omega} = (S_k - S_{k+1}) / \int d(\hbar\omega) f_k(\hbar\omega)$, where $S_k$ is the signal transmitted by filter $k$. The horizontal error bars correspond to the FWHM of the $f_k$ distribution.

**References**

1. Bostedt, C. *et al.* Linac Coherent Light Source: The first five years. *Reviews of Modern Physics* **88**, 015007 (2016).
2. Bilderback, D.H., Elleaume, P. & Weckert, E. Review of third and next generation synchrotron light sources. *J Phys B-at Mol Opt* **38**, S773-S797 (2005).
3. Altarelli, M. The European X-ray Free-Electron Laser:toward an ultra-bright, high repetition-rate x-ray source. *High Power Laser Science and Engineering* **3**, e18 (2015).
4. Esarey, E., Schroeder, C.B. & Leemans, W.P. Physics of laser-driven plasma-based electron accelerators. *Rev. Mod. Phys.* **81**, 1229-1285 (2009).
5. Kneip, S. *et al.* Near-GeV acceleration of electrons by a nonlinear plasma wave driven by a self-guided laser pulse. *Phys. Rev. Lett.* **103**, 035002 (2009).
6. Clayton, C.E. *et al.* Self-guided laser wakefield acceleration beyond 1 GeV using ionization-induced injection. *Phys. Rev. Lett.* **105**, 105003 (2010).
7. Liu, J.S. *et al.* All-optical cascaded laser wakefield accelerator using ionization-induced injection. *Phys. Rev. Lett.* **107**, 035001 (2011).
8. Kim, H.T. *et al.* Enhancement of electron energy to the multi-GeV regime by a dual-stage laser-wakefield accelerator pumped by petawatt laser pulses. *Phys. Rev. Lett* **111**, 165002 (2013).
9. Wang, W.T. *et al.* Control of seeding phase for a cascaded laser wakefield accelerator with gradient injection. *Appl Phys Lett* **103**, 243501 (2013).
10. Wang, X. *et al.* Quasi-monoenergetic laser-plasma acceleration of electrons to 2 GeV. *Nat. Commun.* **4**, 1988 (2013).
11. Leemans, W.P. *et al.* Multi-GeV electron beams from capillary-discharge-guided subpetawatt laser pulses in the self-trapping regime. *Phys. Rev. Lett.* **113**, 245002 (2014).
12. Corde, S. *et al.* Femtosecond x rays from laser-plasma accelerators. *Rev. Mod. Phys.* **85**, 1-48 (2013).
13. Esarey, E., Shadwick, B.A., Catravas, P. & Leemans, W.P. Synchrotron radiation from electron beams in plasma-focusing channels. *Physical Review E* **65** (2002).
14. Kostyukov, I., Kiselev, S. & Pukhov, A. X-ray generation in an ion channel. *Physics of Plasmas* **10**, 4818 (2003).
15. Rousse, A. *et al.* Production of a keV X-Ray Beam from Synchrotron Radiation in Relativistic Laser-Plasma Interaction. *Phys. Rev. Lett.* **93** (2004).
16. Albert, F. *et al.* Betatron oscillations of electrons accelerated in laser wakefields characterized by spectral x-ray analysis. *Physical Review E* **77** (2008).
17. Kiselev, S., Pukhov, A. & Kostyukov, I. X-ray generation in strongly nonlinear plasma waves. *Phys Rev Lett* **93**, 135004 (2004).
18. Cipiccia, S. *et al.* Gamma-rays from harmonically resonant betatron oscillations in a plasma wake. *Nat. Phys.* **7**, 867-871 (2011).
19. Popp, A. *et al.* All-Optical Steering of Laser-Wakefield-Accelerated Electron Beams. *Phys.*

*Rev. Lett.* **105** (2010).

20. Ta Phuoc, K. *et al.* Betatron radiation from density tailored plasmas. *Physics of Plasmas* **15**, 063102 (2008).
21. Chen, L.M. *et al.* Bright betatron X-ray radiation from a laser-driven-clustering gas target. *Sci. Rep.* **3** (2013).
22. Yan, W. *et al.* Concurrence of monoenergetic electron beams and bright X-rays from an evolving laser-plasma bubble. *Proceedings of the National Academy of Sciences of the United States of America* **111**, 5825-5830 (2014).
23. Rykovanov, S.G., Schroeder, C.B., Esarey, E., Geddes, C.G.R. & Leemans, W.P. Plasma Undulator Based on Laser Excitation of Wakefields in a Plasma Channel. *Phys. Rev. Lett.* **114** (2015).
24. Chen, M. *et al.* Tunable synchrotron-like radiation from centimeter scale plasma channels. *Light: Science & Applications* **5**, e16015 (2016).
25. Yi Xu, J.L., Wenkai Li, Fenxiang Wu, Yanyan Li, Cheng Wang, Zhaoyang Li, Xiaoming Lu,Yanqi Liu, Yuxin Leng, Ruxin Li, Zhizhan Xu A Stable 200TW / 1Hz Ti:sapphire laser for driving full coherent XFEL. *Optics & Laser Technology* **79**, 141 (2016).
26. Ferri, J. *et al.* Effect of experimental laser imperfections on laser wakefield acceleration and betatron source. *Scientific reports* **6**, 27846 (2016).
27. Li, W.T. *et al.* Observation of laser multiple filamentation process and multiple electron beams acceleration in a laser wakefield accelerator. *Physics of Plasmas* **20**, 113106 (2013).
28. Changhai Yu, R.Q., Wentao Wang, Jiansheng Liu, Wentao Li, Cheng Wang, Zhijun Zhang, Jiaqi Liu, Zhiyong Qin, Ming Fang, Ke Feng, Ying Wu, Ye Tian, Yi Xu, Fenxiang Wu, Yuxin Leng, Xiufeng Weng, Jihu Wang, Fuli Wei, Yicheng Yi, Zhaohui Song, Ruxin Li& Zhizhan Xu Ultrahigh brilliance quasimonochromatic MeV γ-rays based on self-synchronized all-optical Compton scattering. *Sci. Rep.*, 29518 (2016).
29. W. T. Wang, W.T.L., J. S. Liu, Z. J. Zhang, R. Qi, C. H. Yu, J. Q. Liu, M. Fang, Z. Y. Qin, C. Wang, Y. X u , F. X. Wu, Y. X. Leng, R. X. Li and Z. Z. Xu High-Brightness High-Energy Electron Beams from a Laser Wakefield Accelerator via Energy Chirp Control. *Phys. Rev. Lett.* **117**, 124801 (2016).
30. Dasilva, L.B. *et al.* Electron-Density Measurements of High-Density Plasmas Using Soft-X-Ray Laser Interferometry. *Phys. Rev. Lett.* **74**, 3991-3994 (1995).
31. Cauble, R. *et al.* Micron-resolution radiography of laser-accelerated and X-ray heated foils with an X-ray laser. *Phys Rev Lett* **74**, 3816-3819 (1995).
32. Esarey, E., Sprangle, P., Krall, J. & Ting, A. Overview of plasma-based accelerator concepts. *Ieee T Plasma Sci* **24**, 252-288 (1996).
33. Monot, P. *et al.* Experimental demonstration of relativistic self-channeling of a multiterawatt laser pulse in an underdense plasma. *Phys. Rev. Lett.* **74**, 2953-2956 (1995).
34. Muggli, P. *et al.* Boundary effects. Refraction of a particle beam. *Nature* **411**, 43 (2001).
35. Katsouleas, T. *et al.* Laser steering of particle beams: refraction and reflection of particle beams. *Nucl Instrum Meth A* **455**, 161-165 (2000).
36. Buck, A. *et al.* Shock-front injector for high-quality laser-plasma acceleration. *Phys. Rev. Lett.* **110**, 185006 (2013).
37. Tsung, F.S. *et al.* Near-GeV-energy laser-wakefield acceleration of self-injected electrons in a centimeter-scale plasma channel. *Phys Rev Lett* **93**, 185002 (2004).

38. Kalmykov, S.Y. *et al.* Electron self-injection into an evolving plasma bubble: Quasi-monoenergetic laser-plasma acceleration in the blowout regime. *Physics of Plasmas* **18** (2011).
39. Thorn, D.B. *et al.* Spectroscopy of betatron radiation emitted from laser-produced wakefield accelerated electrons. *Rev Sci Instrum* **81**, 10E325 (2010).
40. Fourmaux, S. *et al.* Demonstration of the synchrotron-type spectrum of laser-produced Betatron radiation. *New Journal of Physics* **13**, 033017 (2011).
41. Plateau, G.R. *et al.* Low-Emittance Electron Bunches from a Laser-Plasma Accelerator Measured using Single-Shot X-Ray Spectroscopy. *Phys. Rev. Lett.* **109** (2012).
42. Ta Phuoc, K. *et al.* All-optical compton gamma-ray source. *Nat. Photonics* **6**, 308-311 (2012).
43. Powers, N.D. *et al.* Quasi-monoenergetic and tunable X-rays from a laser-driven compton light source. *Nat. Photonics* **8**, 28-31 (2013).
44. Schmid, K. *et al.* Density-transition based electron injector for laser driven wakefield accelerators. *Phys Rev Spec Top-Ac* **13**, 091301 (2010).
45. Fourment, C. *et al.* Broadband, high dynamics and high resolution charge coupled device-based spectrometer in dynamic mode for multi-keV repetitive x-ray sources. *Rev. Sci. Instrum.* **80**, 083505 (2009).

**Acknowledgements**

This work was supported by the National Natural Science Foundation of China (Grant No. 11425418, 11127901, 61521093, and 11505263) and the Strategic Priority Research Program (B) (Grant No. XDB16), State Key Laboratory Program of Chinese Ministry of Science and Technology.

**Author contributions**

All authors contributed significantly to the work presented in this paper.

**Competing financial interests**

The authors declare no competing financial interests.

**Correspondence**

Correspondence and requests for materials should be addressed to Jiansheng Liu (e-mail: *michaeljs_liu@siom.ac.cn*), R.L. (*ruxinli@mail.shcnc.ac.cn*) and Z.X. (*zzxu@mail.shcnc.ac.cn*)

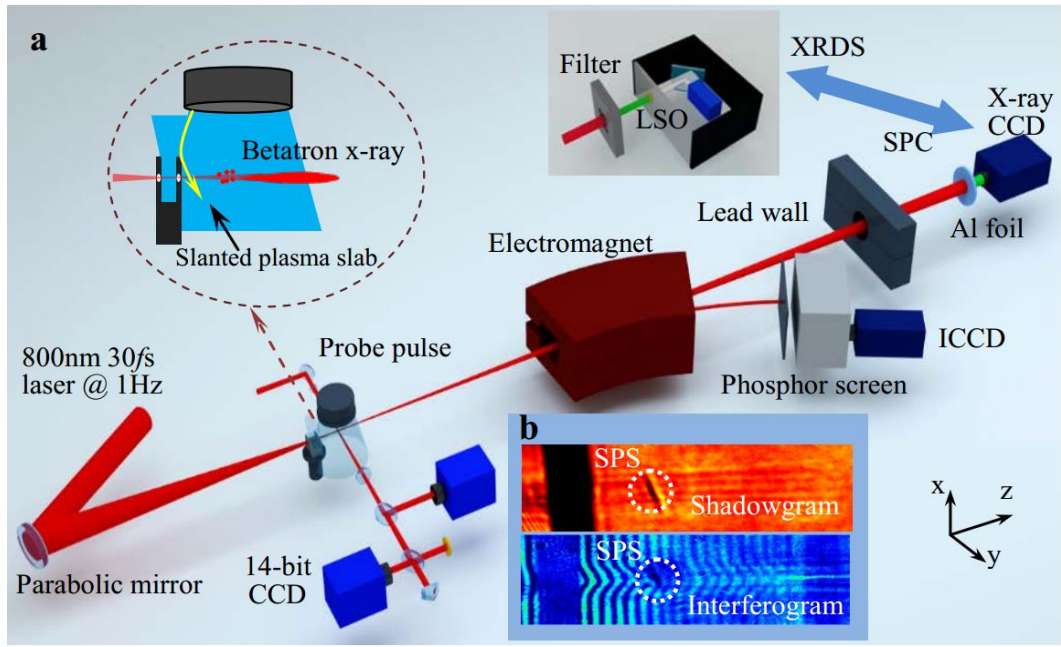

**Figure 1 | Layout of the enhanced betatron radiation experimental setup**. (**a**)The 0.8-um-wavelength, 33*fs*-duration pulses from the 200 TW Ti:sappire laser system were focused onto the two-segment (0.8+ 3 mm) pure helium gas jets onto the supersonic nozzles. About 5% of the pump laser was split off as a probe pulse, which crossed the interaction region perpendicularly into two CCD cameras to measure the plasma density profile and overview the laser evolution. The produced e-beams are deflected by a 90-cm-long dipole electromagnet with a maximum magnetic field of 1.1 Tesla, and measured by a Lanex phosphor screen (PS) imaged onto an intensified charge-couple device 16-bit (ICCD) in a single shot. The betatron radiation x-rays was detected by an x-ray CCD or an x-ray spectra analyzer using a transmission filter and a $Lu_2SiO_4$(LSO)-crystal scintillator, which were placed downstream 4.7 m or further away from the LWFA; (**b**) The measured SPS via an optical interferometer and a shawdowgraphy.

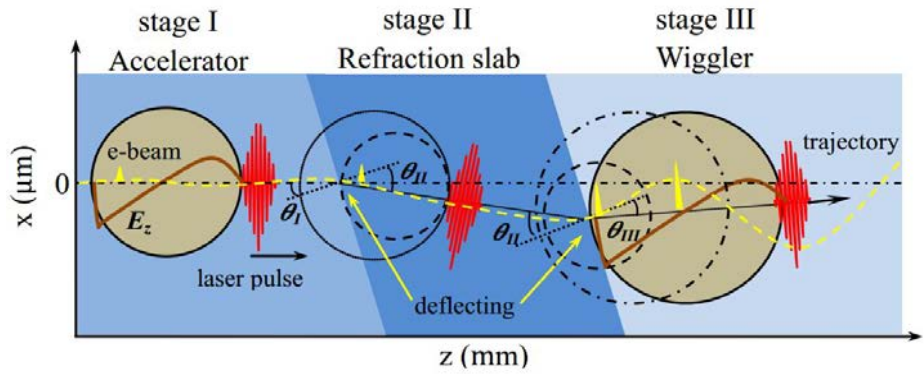

**Figure 2 | Principle of steering electron beam in a wakefield.** The slanted plasma slab (SPS) was inserted between the accelerator (stage I) and the plasma wiggler (stage III). Laser-driven wakefield was deflected due to the refraction at the boundaries (I→II and II→III) of the SPS, resulting in a misaligned propagation of the laser and the accelerated electron beam. The trapped electrons can thus obtain a much larger transverse oscillation amplitude in the deflected plasma wiggler and the peak energy of the low-energy-spread electron beam varies little for its quick phase slippage to the middle of the wake in the plasma wiggler regime.

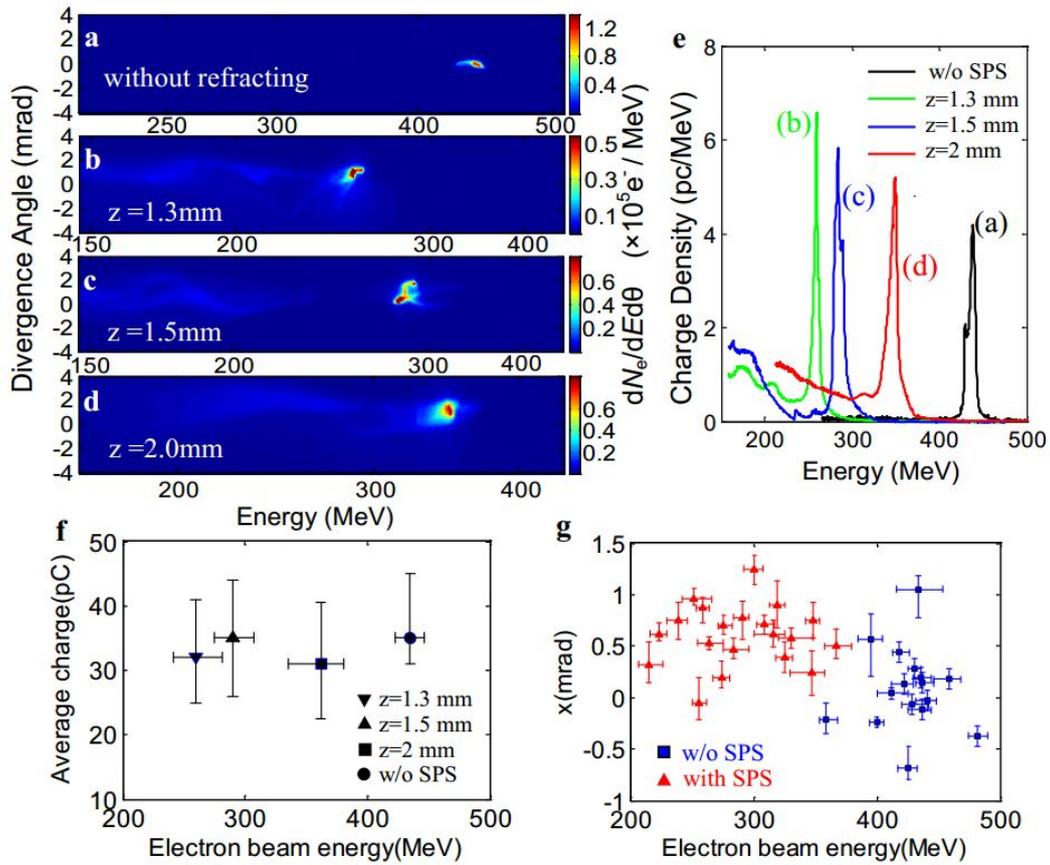

**Figure 3 | Electron beam measurement. (a-d)** Single-shot typical electron spectra without and with introducing the SPS, z is the SPS position. **(e)** The peak energy, FWHM energy spread, integrated charge in the peak and rms divergence of the corresponding *e* beams are: 438MeV, 3.0%, 46pC, 0.24mrad in **a**; 259MeV, 1.8%, 31pC, 1.16mrad for z=1.3mm in **b**; 284MeV, 3.5%, 56pC, 0.94mrad for z=1.5mm in **c**; 351MeV, 2.6%, 45pC, 1.1mrad for z=2.0mm in **d**, respectively. **(f)** Measured average *e*-beam charge in the peak from a series shots over 50 for each case. The horizontal and vertical error bars correspond to the uncertainties of the *e*-beam peak energy and charge. **(g)** Each dot represents the *e*-beam average central position over five shots without (blue squares) and with (red triangles) the SPS at different z, and the vertical error bars indicate the statistic fluctuation of *e*-beam central positions.

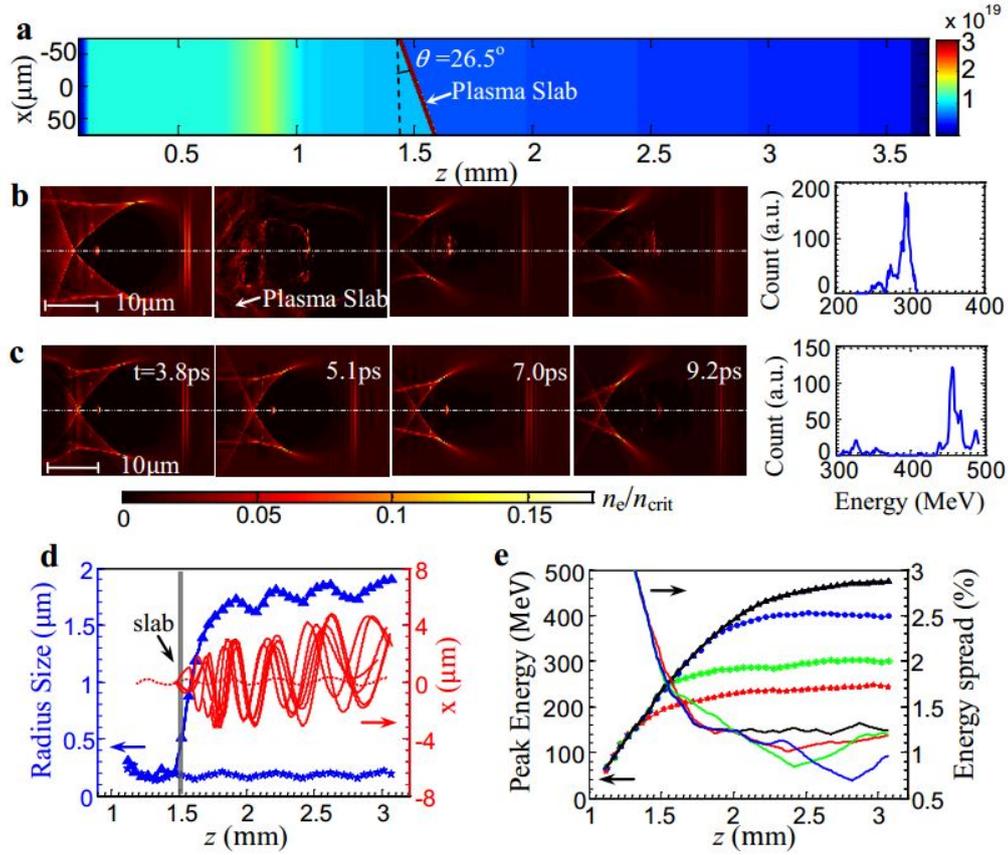

**Figure 4 | PIC simulation results. (a)** The plasma density distribution with the SPS; **(b-c)** Snapshots of the 2-D electron density distribution and wakefield structures for the cases with and without SPS from the PIC simulations. The white dashed line represent the incident direction of the laser pulse. Both the electron spectra at t=9.2 ps were plotted; **(d)** Transverse size for the wiggling e-beam without (blue star) and with (blue triangle) the SPS. Trajectories of accelerated electrons with (red solid line) and without (red dashed line) the SPS were tracked; **(e)** Evolution of the peak energy and rms energy spread of the e-beams without (black) and with the SPS placed at z=1.3 mm (red), 1.5 mm (green) and 2 mm (blue), respectively.

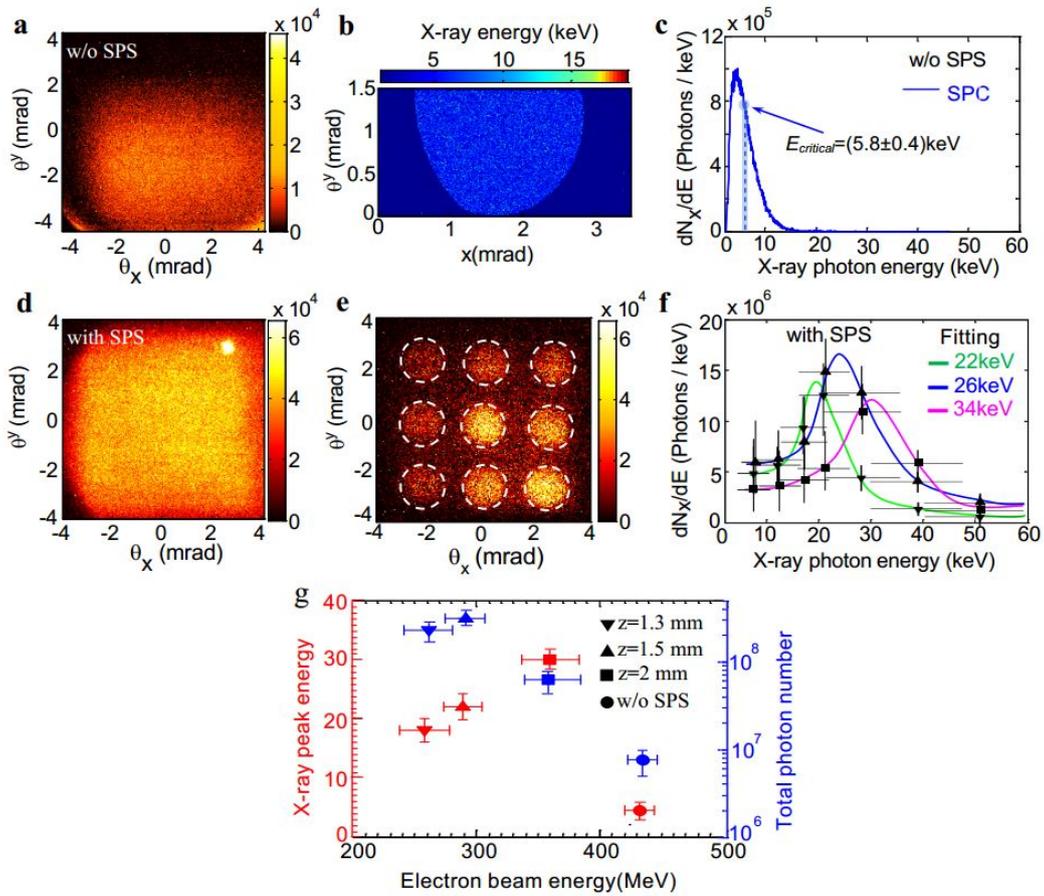

**Figure 5 | X-ray spectra measurement. (a-c)** In the case of no SPS, typical beam patterns recorded by the XRDS, background-subtracted x-ray beam recorded on the x-ray CCD in the SPC mode and measured radiation spectra for the corresponding *e* beams in Fig. 3 **a**, respectively. **(d-e)** Typical enhanced betatron x-ray beam patterns without the filter, transmitted x-ray beam profile through a 3×3 circle-grid filter packed with 0.1-5mm thick attenuated sheets made of Al, Cu, Pb materials. **(f)** Measured x-ray spectra with the SPS (z=1.3, 1.5, 2mm) for the corresponding *e* beams in Fig. 3**b-d**. Data points were obtained by analyzing the measured transmittance of the filter. Horizontal and vertical error bars represent the FWHM of the $f_k$ distribution (see Methods) and the measurement uncertainties, respectively. **(g)** Measured average x-ray peak energy and total photon number from a series shots for each case, plotted versus with the *e*-beam peak energy.